\documentclass[superscriptaddress,dvipsnames,nofootinbib,amsmath,amssymb,aps,prd,reprint]{revtex4-2}

\usepackage[titletoc,toc,title]{appendix}

\usepackage{mathtools}
\usepackage{amsmath, amsfonts,bm}
\usepackage[utf8]{inputenc}

\usepackage{xcolor}
\usepackage{graphicx}
\usepackage{float}
\usepackage{subfig}

\usepackage[a4paper, left=20mm,right=20mm,top=20mm,bottom=20mm]{geometry}
\usepackage[colorlinks=true,hyperfootnotes=true, linkcolor=RedViolet, citecolor=BlueViolet, urlcolor=BlueViolet]{hyperref}

\newcommand\mathcomma{\,,}
\newcommand\mathperiod{\,.}

\DeclareMathAlphabet{\mathup}{OT1}{\familydefault}{m}{n}

\def\dd{\mathrm{d}}

\renewcommand\vec[1]{\bm{#1}}

\newcommand{\be}{\begin{equation}} 
\newcommand{\ee}{\end{equation}}

\usepackage{array}
\newcommand{\PreserveBackslash}[1]{\let\temp=\\#1\let\\=\temp}
\newcolumntype{C}[1]{>{\PreserveBackslash\centering}p{#1}}
\newcolumntype{R}[1]{>{\PreserveBackslash\raggedleft}p{#1}}
\newcolumntype{L}[1]{>{\PreserveBackslash\raggedright}p{#1}}

\begin{document}

\title{Scalar field dark matter and dark energy: A hybrid model for the dark sector}
\author{Carsten van de Bruck}
\author{Gaspard Poulot}
\author{Elsa M. Teixeira}
\affiliation{School of Mathematics and Statistics, University of Sheffield, Hounsfield Road, Sheffield S3 7RH, United Kingdom}

\begin{abstract}
Diverse cosmological and astrophysical observations strongly hint at the presence of dark matter and dark energy in the Universe. One of the main goals of Cosmology is to explain the nature of these two components. It may well be that both dark matter and dark energy have a common origin. In this paper, we develop a model in which the dark sector arises due to an interplay between two interacting scalar fields. Employing a hybrid inflation potential, we show that the model can be described as a system of a pressureless fluid coupled to a light scalar field. We discuss this setup's cosmological consequences and the observational signatures in the cosmic microwave background radiation and the large-scale structures. 
\end{abstract}

\maketitle

{\hypersetup{linkcolor=black}
}

\section{Introduction}

The standard model of Cosmology, referred to as the $\Lambda$CDM model, successfully explains various observations about our Universe. It makes numerous non--trivial predictions that have been confirmed by cosmological data, such as the spectrum of anisotropies in the cosmic microwave background (CMB) radiation and the overall shape of the matter power spectrum (see, \textit{e.g.} \cite{1804633} for a review, historical overview and references). According to the picture developed in the last decades, structure formation is achieved \textit{via} gravitational instability resulting in the smallest collapsed systems forming first. One essential ingredient required to explain the distribution of structures in the Universe is cold dark matter (CDM), which must interact with standard model particles only very weakly (if at all). Another component needed in the $\Lambda$CDM model is the cosmological constant $\Lambda$, which accounts for the observed late-time accelerated expansion of the Universe. It is broadly categorised as dark energy (DE), subject to the condition that its value must be minimal compared to typical energy scales in Particle Physics. Therefore, alternative explanations for the accelerated expansion have been proposed, such as scalar fields. These allow for the negative pressure needed to explain the accelerated expansion in Einstein's theory. Other attempts to extend the $\Lambda$CDM model include modifications to general relativity; see, \textit{e.g.} \cite{Clifton:2011jh} and \cite{Joyce:2016vqv} for references and overviews. 

Dark matter (DM) and DE make up the so-called dark sector. It is a major goal of cosmology to illuminate their properties and origins. There is a plethora of phenomenological proposals for DM, motivated by theories beyond the standard model of particle physics. Candidates for DM range from weakly interacting massive particles to light scalar fields; see, \textit{e.g.}  \cite{Jungman:1995df,Bertone:2004pz,Holman:1982tb,Blumenthal:1984bp,KolbWimplzilla1,Hlozek:2014lca,Marsh:2015xka,Ballesteros:2016euj,Hlozek:2017zzf,Roszkowski:2017nbc} and references therein. DE is often seen as a separate issue, not related to DM. In this work, we propose a shared origin for the dark sector as two scalar fields driven by a shared potential energy $V$\footnote{Models in which DM and DE are both scalar fields have been considered e.g. in \cite{DAmico:2016jbm,CarrilloGonzalez:2017cll,Benisty:2018qed,Brandenberger:2019jfh,Brandenberger:2020gaz,Johnson:2020gzn,Sa:2021eft,Johnson:2021wou}.}. Our choice for the potential energy is the one used in hybrid inflation \cite{Lindehybridinflation}, allowing for a hierarchy of masses for DM and DE, which we discuss in more detail in the next Section. The DM field is identified with the heavier field, and its mass is set by the expectation value of the DE field, corresponding to the flat direction of the potential. We will show that the DE field is limited to changing very slowly under these conditions. One consequence of the theory proposed is that the current period of accelerated expansion is transient. In the future, both fields will settle at the actual minimum, for which the potential energy vanishes. The subsequent evolution of the Universe is then determined by other factors, such as the curvature of space. 

The paper is organised as follows: In Section~\ref{sec:model}, we present the model. The conditions on the model parameter are discussed in Section~\ref{sec:param}. To study the cosmological background dynamics and calculate the evolution of perturbations, we develop a fluid description for the DM field in Section~\ref{sec:fluid}. In Section~\ref{sec:eqs}, we describe the evolution of the Universe and the predictions for the CMB anisotropies and large-scale structures spectra. The results and phenomenology of the model are discussed in Section~\ref{sec:results}. We conclude in Section~\ref{sec:conc}.

\section{Model} \label{sec:model}

In this section, we discuss the field contents of the model studied in this paper. The set-up under consideration is based on that of hybrid inflation  \cite{Lindehybridinflation} with the addition of the standard model fields and is defined through the following action: 

\begin{eqnarray}
     S &=&\int d^4x \sqrt{-g} \left[ \frac{1}{2} \text{M}^2_{\text{Pl}} R-\frac{1}{2}(\partial\phi)^2-\frac{1}{2}(\partial\chi)^2-V(\phi,\chi)\right]\nonumber \\
     &+& S_{\rm SM}  \mathperiod
\end{eqnarray}
We aim to have $\phi$ playing the role of DE and $\chi$ being DM. The standard model fields are accounted for in the action $S_{\rm SM}$. The term $V(\phi,\chi)$ stands for the effective interaction potential, given in analogy to the one in hybrid inflation as 
\begin{align}
     V(\phi,\chi) &=\frac{\lambda}{4} (M^2-\chi^2)^2+\frac{1}{2}g^2\phi^2\chi^2+\frac{1}{2}{\mu}^2\phi^2 \label{eq:lagrangian}\\ 
     & \equiv  V_0-\frac{1}{2} \lambda M^2 \chi^2+\frac{1}{4}\lambda \chi^4+\frac{1}{2}g^2\phi^2\chi^2+\frac{1}{2}{\mu}^2\phi^2\label{potential2} \mathcomma
\end{align}
where $M$ and $\mu$ are mass scales, $g$ and $\lambda$ are dimensionless coupling constants and $V_0\equiv \frac{1}{4} \lambda M^4$ is the scale of the potential. For $\phi$ and $\chi$ to play the role of DE and DM, respectively, appropriate choices of the parameter values have to be made. We will discuss this in-depth in the next section. The global minimum of the potential is at $\chi = \pm M$ and $\phi=0$, for which the potential energy vanishes. Note that our set-up differs from \cite{Axenides:2004kb}, where $\phi$ and $\chi$ play the role of DM and DE, respectively. Consequently, the physical interpretation of the fields, their dynamics and the choices of parameters change significantly. 

The effective masses of the DM scalar field $\chi$ and of the DE scalar field $\phi$ are determined by the second order derivatives of the potential, given by

\begin{equation}
    m^2_{\chi}\equiv \frac{\partial^2 V}{\partial \chi ^2} = g^2\phi^2-\lambda M^2  + 3\lambda \chi^2 \mathcomma
\label{eq:DM_mass}
\end{equation}
and
\begin{equation}
\label{eq:DE_mass}
    m^2_{\phi} \equiv \frac{\partial^2 V}{\partial \phi^2} = g^2 \chi^2 + {\mu}^2 \mathcomma
\end{equation}
respectively.

We consider a spatially flat Universe described by the Friedmann--Lema\^{i}tre--Robertson--Walker (FLRW) metric with the line--element $$\dd s^2 = - \dd t^2 + a^2(t) \delta_{ij}\dd x^i \dd x^j,$$ where $a(t)$ is the scale--factor. In the following, $H=\dot a/a$ stands for the expansion rate, and over-dots represent derivatives with respect to cosmic time $t$. The equations of motion for each scalar field read
\begin{align} \label{eq:KGphi}
    \ddot{\phi}+3H\dot{\phi}&= -(g^2\chi^2+\mu^2)\phi \mathcomma\\
    \ddot{\chi}+3H\dot{\chi}& =-\lambda \chi^3 + (\lambda M^2 -g^2\phi^2)\chi \mathcomma \label{eq:KGchi}
\end{align}
and the Friedmann equations are 
\begin{align}
    \dot{H} & =-\frac{1}{2 \text{M}^2_{\text{Pl}}}(\rho +P) \mathcomma\\
    H^2 & = \frac{1}{3 \text{M}^2_{\text{Pl}}} \rho \mathcomma
\end{align}
where $\rho$ and $P$ are the collective energy density and pressure of both scalar fields, baryons, and radiation expressed as
\begin{align}
    \rho & =\frac{1}{2} \dot{\phi}^2 + \frac{1}{2} \dot{\chi}^2+V(\phi,\chi)  + \rho_b + \rho_\gamma \mathcomma\\
    P & =\frac{1}{2} \dot{\phi}^2 + \frac{1}{2} \dot{\chi}^2-V(\phi,\chi)+p_\gamma \mathcomma
\end{align}
respectively. For convenience, we split the energy density contributions of each scalar field into two different quantities: 
\begin{align}\label{rhochi}
    \rho_{\chi}&= \frac{1}{2}\dot{\chi}^2-\frac{1}{2} \lambda M^2 \chi^2+\frac{1}{4}\lambda \chi^4+\frac{1}{2}g^2\phi^2\chi^2,\\
    \rho_{\phi}&=\frac{1}{2}\dot{\phi}^2+V_0+\frac{1}{2}\mu^2\phi^2 \mathperiod \label{rhophi}
\end{align}
It is important to emphasise that this splitting is a choice and has no impact on the underlying physics so long as the sum of both parts is equal to the whole energy density of the scalar fields system. The particular choice in Eq.~\eqref{rhochi} is such that all oscillating terms (\textit{i.e.} those containing $\chi$) are grouped to obtain the effective pressureless behaviour needed for structure formation in the matter domination era. The $\phi$-field is expected to behave like a cosmological constant at late times. Nevertheless, we will see that the evolution may still differ at early times, driven by the interaction with the $\chi$-field.

Under this scenario, if $\phi$ is displaced sufficiently far from the origin, then $\chi$ will oscillate around zero. There is an abrupt transition in the shape of the potential when the effective DM mass, given by Eq.~\eqref{eq:DM_mass}, changes from positive to negative. Using Eq.~\eqref{eq:DM_mass} and assuming that $\chi$ is as it oscillates around 0, this happens approximately when $\phi$ reaches a critical value given by
\begin{equation} \label{phicrit}
   \left| \phi_c \right| \approx \frac{\sqrt{\lambda}M}{g} \mathperiod
\end{equation}
For $\phi>\phi_c$, $\chi$ behaves as dark matter, and $\phi$ is a dynamical dark energy component, slowly rolling down its potential. Moreover, the dynamics of $\phi$ is mainly determined by the dominant constant scale in the potential, $V_0$, and the interaction with $\chi$. However, when $\phi$ approaches the critical value $\phi_c$, $\chi$ quickly drops and starts oscillating around $\chi=\pm M$. Simultaneously, $V (\phi, \chi)$ drops to zero leading to a sudden decay of dark energy and implying that the epoch of DE domination is just a transient phenomenon in this theory. 

\section{Conditions on model parameters} \label{sec:param}

In this section, we examine the necessary conditions to fulfil the scenario described above. In other words, we look for constraints on the free parameters $g$, $M$ and $\lambda$. For $\phi$ to play the role of DE, the field needs to roll slowly, and the potential energy needs to be of order $\rho_{\rm DE,0}$, the DE density today. Thus, we demand that $V_0 = \frac{1}{4} \lambda M^4 \approx 10^{-47}$GeV$^{4}$. The contribution from the $\mu^2$--term cannot be larger than this because it also affects the dark energy density. Thus, the mass scale $M$ is of order $10^{-3}$eV, as expected in models with DE. 

On the other hand, for $\chi$ to behave like dark matter, it must oscillate in a quadratic potential from the early Universe onwards \cite{turnerDM}. Firstly to prevent the damping term in Eq.~\eqref{eq:KGchi} from withholding the oscillations, $m_{\chi}\approx g \phi \gg H$ must hold. Secondly, we need to ensure that the quadratic term dominates over the quartic one in Eq.~\eqref{potential2}, which translates into the condition $g^2\phi^2-\lambda M^2 \gg \frac{1}{2}\lambda \chi^2$. As discussed below, $\phi$ does not change significantly during the cosmological evolution, and the value of $\phi$ today must be large, ($\phi_0 \gtrsim 10~\text{M}^2_{\text{Pl}}$). Therefore, the mass of the $\chi$-field, $m_{\chi} = g\phi$, is large unless $g$ is exceedingly small.

At some time $t_i$ in the early universe, $H\approx m_{\chi}$, at which point the field starts to oscillate rapidly around 0 as the expansion rate becomes smaller than the mass. To estimate the temperature of this transition, we assume that the Universe is in the radiation-dominated epoch after an inflationary phase. During this period, $$H^2 = \frac{1}{3\text{M}^2_{\text{Pl}} }\frac{\pi^2}{30}g_*(T) T^4,$$ with $g_*(T)$ being the effective number of relativistic degrees of freedom at a temperature $T$ (which is of the order of several hundred in theories beyond the standard model). Therefore we infer that the oscillations happen at a temperature $$T \approx  10^{15} \left(  \frac{g}{10^{-7}} \right )^{1/2} \left(  \frac{\phi_i}{10 \text{M}^2_{\text{Pl}}} \right )^{1/2}\left(\frac{g_*}{100} \right)^{-1/4} {\rm GeV}.$$ This corroborates the assumption that the field starts to oscillate very early on in the radiation dominated epoch, almost immediately after a period of inflation in this framework. Below we will derive the evolution for the $\chi$--field (eq. (\ref{eq:chisolution})), which allows us to find the initial field amplitude $\chi_i$ in the very early universe. Using the fact that $\rho_{\rm DM,0} \approx g^2 \phi_0^2 \chi_0^2 \approx 4\cdot 10^{-47}$ GeV$^4$ (where the 0 indicates the present time) and that the amplitude evolves as $\chi(t) = \chi_i (a_i/a)^{3/2}=\chi_i (T/T_i)^{3/2}$, we find, using the expression for the temperature above $$\frac{\chi_i}{\rm GeV} \approx 1.4\cdot 10^{6} \left(\frac{g}{10^{-7}} \right)^{-1/4}  \left( \frac{\phi_0}{10 \text{M}^2_{\text{Pl}} } \right)^{-1/4} \left( \frac{g_*}{100}\right)^{-3/8}.$$ This is the initial field amplitude the $\chi$-field must have after inflation in order to predict the right amount of DM today (emphasising again that we assume that the field $\chi$ is responsible for all DM). 

As the model is currently formulated, during inflation the $\phi$--field is light. The only requirement is that its field excursion is large ($\phi \gtrsim 10$ M$_{\rm Pl}$), so that in the radiation dominated epoch the mass of the $\chi$--field also remains large and, as we will see in Section IV, the coupling between $\chi$ and $\phi$ is small enough. Since $\phi$ is light during inflation it is subject to quantum fluctuations, which are of order $H_{\rm inf}/2\pi$, where $H_{\rm inf}$ is the expansion rate during inflation. But the $\phi$--field is a (almost) flat direction and subdominant during the radiation and matter dominated epoch. Therefore the quantum fluctuations will not result in large isocurvature modes in the DE sector. However, the situation with the $\chi$--field is more delicate. If $\phi$ is light during inflation, i.e. $g\phi < H_{\rm inf}$, then the quantum fluctuations of $\chi$ are also of the order $H_{\rm inf}/2\pi$, resulting in potentially large isocurvature modes with an amplitude \cite{Marsh:2014qoa} $$A_I = \frac{(H_{\rm inf}^2/\text{M}^2_{\text{Pl}})}{\pi^2 (\chi_{\rm inf}^2/\text{M}^2_{\text{Pl}})},$$ where $\chi_{\rm inf}$ is the value of $\chi$ during inflation, which has to be of order $10^6$ GeV for $g\approx 10^{-7}$. As it is the case for axion--like fields, there are ways to evade isocurvature bounds. We consider two of these briefly: Firstly, the field $\chi$ is heavy during inflation, so that $g\phi>H_{\rm inf}$. In this case, the isocurvature modes are suppressed. The challenge with this option is that at the end of inflation the field amplitude needs to be large enough so that the $\chi$--field can play the role of DM (or at least be a non--negligible part of the DM sector). Alternatively, the dynamics of $\chi$ during inflation is non--standard, either by coupling $\chi$ to gravity (as in e.g. \cite{Folkerts:2013tua}) or by coupling $\chi$ directly to the inflaton field. In this case, the dark sector is bigger that just the fields $\phi$ and $\chi$ we consider here and it would be interesting to study this option further, also from the model--building perspective. For the rest of the paper, however, we are dealing with the post--inflation period and assume that the isocurvature perturbations can be kept small. 

For the numerical study in the following sections, we select initial conditions, taken at $z_i = 10^{14}$, such that $\phi_i\gg \phi_c$, \textit{i.e.}, $g\phi_i \gg \sqrt{\lambda}M$ from Eq.~\eqref{phicrit} and since $m_\chi\gg H$ we have   
\begin{equation}\label{eq:phiconstraint}
    g\phi_i \gg H \mathcomma
\end{equation}
where a subscript $i$ denotes quantities evaluated at the initial time $z_i$ in the numerical simulations. On the other hand, as previously argued, $\phi$ must be rolling slowly so that $m_{\phi}^2 \ll H^2$. Assuming that $\mu$ is small compared to $g\chi$, the following constraint is obtained from Eq.~\eqref{eq:DE_mass}: 
\begin{equation} \label{eq:chiconstraint}
    g^2 \chi^2 \ll H^2 \mathperiod
\end{equation}
During the matter-dominated epoch, the $\chi$-field dominates the dynamics. As we will describe in more detail in the next section, the DM fractional energy density will eventually start to decrease. At the same time, the Universe keeps expanding, and the $\phi$-field keeps slowly varying until, finally, DE dominates the evolution, driven by the potential energy. Therefore, at early times we require 
\begin{equation}
  \frac{1}{2}\mu^2 \phi^2 + V_0 \ll \frac{1}{2} g^2 \phi^2\chi^2 \mathcomma
\end{equation}
warranting a period of matter domination. From this, we infer 
\begin{align}
    \rho_{\chi}&= \frac{1}{2}\dot{\chi}^2+\frac{1}{2}m^2_{\chi}\chi^2 \nonumber \\
    &\approx m^2_{\chi} \chi^2 \mathcomma
    \label{eq:rho_chi}
\end{align}
where we have used that $m_{\chi}\approx g \phi$ and relied on the fact that the rapidly oscillating $\chi$-field is approximately pressureless when averaged over several oscillation periods. Note that because of how slowly $\phi$ is evolving, the effective DM mass is nearly constant. Solving for $\chi^2$ and replacing in Eq.~\eqref{eq:chiconstraint} leads to
\begin{equation} \label{eq:condition}
    g^2 \frac{\rho_{\chi}}{m^2_{\chi}} \ll H^2 \mathperiod
\end{equation}
During the matter-dominated era, when $\chi$ stands as the predominant contribution, the Friedmann equation can be approximated as
\begin{equation}
    H^2\approx \frac{\rho_{\chi}}{3 \text{M}^2_{\text{Pl}}}.
\end{equation}
Therefore, from Eq.~\eqref{eq:condition}, we arrive at the following condition
\begin{equation} \label{eq:phiconstraint2}
     1\ll \frac{1}{3} \left(\frac{\phi}{{\rm M}_{\rm Pl}} \right)^2 \mathcomma
\end{equation}
where we have employed $m_{\chi}\approx g \phi$. The inequality above can only be satisfied if the $\phi$-field is trans-Planckian, that is, if $\phi \gg {\rm M}_{\rm Pl}$. Therefore, $\phi$ must be at least of the order of the Planck scale to satisfy the constraint in Eq.~\eqref{eq:phiconstraint2} and fulfil the scenario intended in this theory. Consequently, unless $g$ is exceedingly small, this results in a considerably large DM mass, in direct contrast with models with ultralight and light scalar fields as DM candidates \cite{Hu:2000ke,Hui:2016ltb,Hlozek:2014lca}. We remark that this can potentially be accommodated in the WIMPzilla scenario, proposed and analysed in ~\cite{KolbWimplzilla1} and \cite{Kolbwimpzilla2}. 

Another necessary condition for $\chi$ to stand as a viable DM candidate is that the scalar field must remain stable. From a phenomenological perspective, even if kinematically allowed, the decay channel of $\chi$ into $\phi$ is effectively negligible due to the mass scale difference between the fields. Hence we must ensure that the decay rate, given by \cite{Kofman:1997yn}
\begin{equation}
    \Gamma (\chi \chi \rightarrow \phi \phi) = \frac{g^4 \left\langle \chi^2 \right\rangle }{8 \pi m_{\chi}} \mathcomma
\end{equation}
is smaller than the Hubble expansion rate, \textit{i.e.} $\Gamma < H$. We resort to the bracket notation $\langle \cdot \rangle$ to denote averages over one oscillation cycle.
As will be discussed in the next section, the direct dependence of the amplitude on the $\chi$-field and the very slowly evolving $\phi$-field implies that $\Gamma \propto a^{-3}$ (regardless of the epoch), whereas $H$ scales as $a^{-3/2}$ and $a^{-2}$ during matter- and radiation-dominated eras, respectively. These relations imply that the decay width drops much more rapidly than the expansion rate over the Universe's history, ensuring the stability of the DM field for any sensible values of $g<1$.

A final remark before we discuss the cosmological evolution of the system is whether quantum corrections to the potential can spoil the considerations above. In general, quantum corrections to the tree--level potential are expected to be of order ${\rm M}_{\rm Pl}^2$ (choosing ${\rm M}_{\rm Pl}$ to be the natural cut--off). In supersymmetric theories, however, the corrections are of order $\ln(\phi/{\rm M}_{\rm Pl})$ \cite{Lyth:1998xn}. These $\ln$--corrections can be kept small if the coupling constants are small, which is a natural case for the model considered here. We therefore conclude that the model presented here suffers from the same problems as other models of this kind. 

\section{Fluid approximation and dynamics} \label{sec:fluid}

Since the oscillations in the $ \chi$-field are computationally expensive, we wish to find reasonable approximations allowing for the study of the cosmological evolution. More precisely, we recast our framework as an interacting quintessence model through a fluid description of the DM field $\chi$. Despite its similarities with other scalar-field models of DM, one crucial difference in this scenario is that the mass of the DM field evolves as the DE field is slowly rolling. According to previous studies on the cosmological evolution of a scalar field oscillating in a quadratic potential \cite{turnerDM}, it is a well-known result that one can describe its dynamics according to an oscillating envelope with amplitude $\mathcal{A}(t)\propto a^{-\frac{3}{2}}$. Employing the WKB approximation, we can solve for the dynamics of the oscillating scalar field $\chi$ using the conditions derived in the previous section ($g\phi\gg H$ and $\dot\phi/\phi \ll 1$). We arrive at a solution of the form 
\begin{equation} \label{eq:chisolution}
   \chi(t) = \chi_{ i} \left( \frac{\phi_{ i}}{\phi} \right)^{1/2} \left( \frac{a_i}{a} \right)^{3/2} \sin\big(g\phi \left(t-t_{i}\right)\big) \mathperiod
\end{equation}
 where $\chi_{\rm i}$ is the initial amplitude of $\chi$.  Since $\phi$ is evolving slowly, the ratio $\phi_{i} /\phi$ is practically constant, meaning that the $\chi$-field behaves like pressureless dust according to $\rho_\chi \propto \chi^2 \propto a^{-3}$.  Taking the expression for $\rho_{\chi}$ in Eq.~\eqref{eq:rho_chi}, we gather that the energy density of DM averaged over an oscillation period is roughly given by
\begin{equation} \label{eq:fluidDMequation}
    \langle \rho_{\chi} \rangle \approx \rho_{\chi,i}\left( \frac{\phi}{\phi_i}\right) \left( \frac{a_i}{a} \right)^3 \mathcomma
\end{equation}
where $\rho_{\chi,i} = \frac{1}{2}g^2 \phi_i^2 \chi_i^2$ is the energy density of $\chi$ at $t =t_i$.

For simplicity, since we will be considering time--scales much larger than the oscillation span, we drop the bracket notation henceforth, and oscillation-averaged quantities will always be implied. It is worth pointing out that the (averaged) density in Eq.~\eqref{eq:fluidDMequation} depends linearly on $\phi$. Therefore, we obtain the following continuity equation for the oscillation-averaged interacting fluid:
\begin{equation}
    \dot{\rho_{\chi}}+3 H \rho_{\chi}= \frac{\dot{\phi}}{\phi} \rho_{\chi} \mathperiod
    \label{eq:coupling}
\end{equation}
In this context, the equation of motion for the DE field is recast as 
\begin{equation} \label{eq:coupledphiKG}
    \ddot \phi + 3 H \dot \phi = - \frac{1}{\phi} \rho_\chi \mathperiod
\end{equation}
The previous equation is entirely equivalent to a continuity equation for DE, assuming a perfect fluid description for the field as well, with $\rho_{\phi} \approx \dot{\phi}^2/2 + V_0$: 
\begin{equation}
    \dot{\rho_{\phi}}+3H(\rho_{\phi}+P_{\phi})=-\frac{\dot{\phi}}{\phi} \rho_{\chi} \mathcomma
\end{equation}
following conservation of the total energy density of both $\phi$ and $\chi$, as required by the general covariance of Einstein's equations. According to the approximation in  Eq.~\eqref{eq:fluidDMequation}, Eq.~\eqref{eq:coupledphiKG} becomes  
\begin{equation}\label{eq:slowrollKGphi}
    \frac{1}{a^3}\frac{d}{dt}\left(a^3\dot{\phi}\right)= - \frac{\rho_{\chi,i}}{\phi_i} \left(\frac{a_i}{a} \right)^3 \mathcomma
\end{equation}
which, when integrated with respect to time, yields the following expression for the rate of change of the field, $\dot{\phi}$:
\begin{equation} \label{eq:phidotanalytical}
    \dot{\phi}=\left(\frac{a_i}{a}\right)^3 \left(K_i-\frac{\rho_{\chi,i}}{\phi_i}t \right) \mathcomma
\end{equation}
where $K_i\equiv \dot{\phi_i}+\frac{\rho_{\chi,i}}{\phi_i}t_i$ is an integration constant and $\dot{\phi_i}$ is the initial field velocity, that is when $a=a_i$. Hence, provided that the relation between $a$ and $t$ is known, the behaviour of $\phi$ is fully determined. As verified in the previous section, the fluid approximation yields a suitable description right after inflation since the $\chi$-field starts oscillating around 0 immediately after this period has ended, and radiation becomes the dominant contributor in the Universe at $t=t_i$. Consequently, solving Eq.~\eqref{eq:phidotanalytical} in the radiation-dominated epoch, during which $a(t) \propto t^{1/2}$, we obtain
\begin{equation}\label{eq:raddomphi}
    \phi(t)=\phi_i + C_i-A_i \left(\frac{t}{t_i}\right)^{\frac{1}{2}}- B_i \left(\frac{t}{t_i}\right)^{-\frac{1}{2}}\mathcomma
\end{equation}
where $C_i \equiv 2 \left( \frac{\rho_{\chi,i}}{\phi_i} t_i^2+K_i t_i\right)$, $A_i \equiv 2 \frac{\rho_{\chi,i}}{\phi_i} t_i^{2}$ and $B_i \equiv 2 K_i t_i$ are integration constants. Therefore during this period, the energy density of $\phi$ scales according to
\begin{equation}
    \rho_{\phi} \propto \dot{\phi}^2 \propto a^{-1} \mathperiod
\end{equation}
Eq.~\eqref{eq:raddomphi} sets the field's evolution until the matter-radiation equality at $t_{\rm eq}$.  When the matter-dominated era begins,  $a(t)\propto t^{2/3}$ which leads to the following solution of Eq.~\eqref{eq:phidotanalytical}:
\begin{equation}\label{eq:matterdomphi}
    \phi(t)=\phi_{\rm eq}+ C_{\rm eq} -A_{\rm eq} \ln{\left(\frac{t}{t_{\rm eq}}\right)}-B_{\rm eq}\left( \frac{t}{t_{\rm{eq}}} \right)^{-1} \mathcomma
\end{equation}
where equivalently $C_{\rm eq} \equiv t_{\rm eq} K_{\rm eq} $, $A_{\rm eq} \equiv  t_{\rm eq}^2 \frac{\rho_{\chi,\rm eq}}{\phi_{\rm eq}}$, and  $B_{\rm eq} \equiv t_{\rm eq} K_{\rm eq}$ are constants depending on initial conditions taken at radiation-matter equality, denoted by the subscript ``eq". The constant $K_{\rm eq}$ is defined in analogy to $K_i$ in Eq.~\eqref{eq:phidotanalytical}, with each quantity taken at time $t_{\rm eq}$ instead of $t_{i}$. Since the field is slow-rolling, it is reasonable to assume $\dot{\phi_{\rm eq}} \ll  t_{\rm eq} \rho_{\chi,{\rm eq}}/\phi_{\rm eq}$, resulting in $B_{\rm eq} \approx t_{\rm eq} A_{\rm eq}$. 
It is worth noting that Eq.~\eqref{eq:matterdomphi} implies that $ \dot{\phi} \propto a^{-\frac{3}{2}}$  since $t \propto a^{\frac{3}{2}}$ during matter domination. Moreover, considering that the coupling to DM is the main driver of the field's dynamics and accordingly $\dot\phi^2 \gg V_0$, we arrive at 
\begin{equation}
    \rho_{\phi} \propto \dot{\phi}^2 \propto a^{-3} \mathperiod
    \label{eq:scaling}
\end{equation}
It is noteworthy that, during this regime, the DE component scales with ordinary matter and CDM. This scaling is not a general feature of interacting dark energy models with a constant potential, and we have checked that this is the only formulation leading to this singular behaviour. Solutions of this kind are relevant to address the cosmic coincidence problem of $\Lambda$CDM concerning the comparable magnitude for the energy density of DE ($\Lambda$ in the standard model) and CDM at present \cite{Ferreira:1997hj,Bahamonde:2017ize,vandeBruck:2016jgg,Teixeira:2019tfi}. In the next section, we illustrate and analyse the dynamics in this regime through numerical simulations.

The form of the coupling term on the right-hand side of Eq.~\eqref{eq:coupling} implies that $\phi$ must be large up until the current cosmological era, in line with the discussion in Sec.~\ref{sec:param}, as required to avoid drastic deviations from the $\Lambda$CDM case. Albeit counter-intuitive at first glance, this framework hinges on the fact that $\phi$ is rolling slowly as $\dot{\phi}/(\phi H) \ll 1 $,  which we have also confirmed numerically, implying that $\phi> M_{\rm Pl}$, according to Eq.~\eqref{eq:phiconstraint2}.
 
From a mathematical point of view, the result of the fluid approximation of the system considered here is analogous to encapsulating the effect of a 5th--force, mediated by a dark energy scalar field, in a conformally rescaled metric that determines the geodesics for the dark matter particles $\widetilde{g}_{\mu \nu}$, given by 
\begin{equation}
    \widetilde{g}_{\mu \nu} = C\left( \phi \right) g_{\mu \nu} \mathcomma
\end{equation}
in terms of the gravitational metric $g_{\mu \nu}$, with the conformal factor identified as
\begin{equation}
    C(\phi) = \frac{\phi^2}{M_{\rm Pl}^2} ~~~\mathrm{for}~ |\phi| > |\phi_c| \mathperiod~~~
\end{equation}
The transformation is always invertible as $\lvert \phi \rvert> \lvert \phi_c \rvert $. Note that the fluid approximation will break down well before $\phi$ can approach $0$. Moreover, it follows that $C \left( \phi \right) > 0 $ and the Lorentzian signature of the metric is preserved, avoiding any instabilities related to metric singularities. In this framework, the form of the coupling function for the fluid approximation is recovered and reads:

\begin{equation}
   Q = - \frac{C_{\phi}}{2C} \rho_{\chi} = - \frac{\rho_{\chi}}{\phi} \mathperiod
   \label{eq:qconf}
\end{equation}
By modelling both components of the dark sector as perfect fluids, it is possible to rewrite the relevant dynamical equations, such as the conservation relations. We resort to conformal time in our numerical work, defined by $\dd \tau = \dd t/a$. A prime indicates derivatives with respect to $\tau$, and the Hubble rate is rescaled as $\mathcal{H} = a H$. The equations for the DM and DE fluids read 
\begin{equation}
    \rho'_{\chi} + 3 \mathcal{H} \rho_{\chi} = - Q \phi' = \frac{\phi'}{\phi} \rho_{\chi} \mathcomma
    \label{eq:conschi}
\end{equation}
\begin{equation}
    \rho'_{\phi} + 3 \mathcal{H} \left( \rho_{\phi} + p_{\phi} \right) = Q \phi' = - \frac{\phi'}{\phi} \rho_{\chi} \mathperiod
    \label{eq:consphi}
\end{equation}
These equations lay out the energy exchange between the fluids, with the direction directly related to the sign of $\phi'/\phi$. If the ratio is positive, it is DE sourcing the DM component, while if it is negative, there will be an energy flow from $\phi$ to the dark matter fluid. Regardless of the initial conditions chosen, we find that $\phi$ and $\phi'$ always have opposite signs. Consequently, this model exhibits a unidirectional energy transfer from the $\chi$ fluid to the $\phi$-field. The modified Klein-Gordon equation encodes the same information:

\begin{equation}
   \phi'' + 2 \mathcal{H} \phi' = - \frac{a^2}{\phi} \rho_{\chi} = a^2 Q \mathcomma
   \label{eq:kgphip}
\end{equation}
and can be numerically integrated for different realisations of the system yielding particular solutions for the dynamical evolution of the model.

For numerical purposes, the only free model-specific parameters are the initial conditions for the DE scalar field $\phi_i = \phi \left( \tau_i \right)$ and $\phi'_i = \phi' \left( \tau_i \right)$ and the scale of the hybrid potential $V_0$. It is important to note that the parameters in the potential energy in  Eq.(\ref{eq:lagrangian}) drop out completely from the calculation, meaning we do not need to choose their values to solve the system numerically so long as we assume that they satisfy the constraints derived in Sec.(\ref{sec:param}). However, we set $\mu$ equal to zero for simplicity since it does not contribute up to current times.   Without loss of generality, we compute $V_0$ through a shooting method for the fiducial value of the present DE relative energy density: $\Omega_{\phi}^0 = \rho_{\phi}^0/(3 {\rm M}_{\rm Pl}^2 H_0^2)$, where $H_0$ is the Hubble expansion rate at the present epoch. Moreover, and as previously mentioned, the value of $\phi'_i$ has no impact on the dynamics as the scalar field is quickly driven towards the minimum deep in the radiation-dominated epoch where its contribution is negligible. For this reason, and without loss of generality, in the numerical study, we always take $\phi'_i = 0 $. In this way, the analysis presented can be focused on the effects of varying the only free parameter: the initial condition for the scalar field $\phi_i$. We will focus only on scenarios for which $\phi_i>0$, as the solutions for $\phi_i<0$ would lead to the same dynamics starting from the opposite side of the symmetric potential.

\section{Cosmological perturbations and observables} \label{sec:eqs}

Following the discussion on the background evolution, we now map the cosmological perturbations onto an interacting DE model. We are interested in studying the modifications to the gravitational interaction in contrast to $\Lambda$CDM and assessing the measurable imprints left by the approximations made at the background level. For this purpose, we consider perturbations in the Newtonian gauge \cite{Mukhanov:1990me}, corresponding to the following line element
\begin{equation}
     \dd s^2 = a^2(\tau)\left[ -\left( 1+2\Psi \right)\dd \tau^2 + \left( 1-2\Phi \right)\delta_{ij}\dd x^i \dd x^j \right] \mathcomma
\end{equation}
where $\Psi(\tau,\vec{x})$ and $\Phi(\tau,\vec{x})$ are the conventional Newtonian scalar potentials. For the remainder of this section, a $\delta$ denotes perturbed quantities, and since we are dealing with a system of scalar fields, the anisotropic stress vanishes. Moreover, we work in Fourier space, such that the mapping $\nabla^2 \rightarrow - k^2$ holds for the spatial derivatives of the respective quantities.

The equations of motion for $ \delta \phi$ and  $\delta \rho_{\chi}$ in the coupled DE framework are 
\begin{equation}
    \delta \phi'' + 2 \mathcal{H} \delta \phi' + k^2 \delta \phi = \left( \Psi' + 3 \Phi' \right) \phi' + 2 a^2 Q \Psi + a^2 \delta Q  \mathcomma
\end{equation}
\begin{equation} \label{eq:pertrho}
    \delta_{\chi}'=- (\theta_{\chi}-3 \Phi') +\frac{Q}{\rho_{\chi}}\phi'\delta_{\chi} - \frac{Q}{\rho_{\chi}}\delta \phi'-\frac{\theta'}{\rho_{\chi}}\delta Q \mathcomma
\end{equation}
where we have defined the density contrast $\delta_{\chi} = \delta \rho_{\chi}/\rho_{\chi}$ and the perturbed coupling $\delta Q$ is given by

\begin{equation}
  \delta Q =\frac{\rho_{\chi} \delta \phi - \phi \delta \rho_{\chi}}{\phi^2}  \mathperiod
\end{equation}

It is worth remarking that in Eq.~\eqref{eq:pertrho}, both the equation of state $w_{\chi}=\frac{p_{\chi}}{\rho_{\chi}}$ and the sound speed $c_s^2=\frac{\delta p_{\chi}}{\delta \rho_{\chi}}$ of the DM fluid were set to zero. The former assumption is motivated by the study of the background dynamics in the previous section, while the latter is justified when looking at the explicit form of the sound speed for an oscillating scalar field, which under this approximation becomes \cite{Hlozek:2014lca}, 
\begin{equation}
    c^2_s=\frac{k^2/(4 m_{\chi}^2 a^2)}{1+k^2/(4 m_{\chi}^2 a^2)} \mathperiod
\end{equation}
This equation strictly holds for an uncoupled scalar field, but since the coupling considered here is small, it captures the essential physics. Since $m_{\chi}$ is required to be considerably large in our model and we are considering scales $k \ll 2 m_{\chi} a$, it is a reasonable assumption to take $c_s^2=0$, which we have also confirmed numerically. 

To better appreciate the influence of the coupling on the evolution of the density matter perturbations, we look at scales in the sub-horizon limit ($ k \gg \mathcal{H}$) together with the quasi-static approximation. The latter relies on the matter and field perturbations being the main contributors to the time variation of the gravitational potentials. In practice, this implies neglecting the time derivatives of the perturbations and metric potentials, leading to the following simplification for the equation of motion for $\delta_{\chi}$ (neglecting the contribution of baryons) \cite{vandeBruck:2015ida,vandeBruck:2020fjo}:

\begin{equation}
\delta_{\chi}'' + \mathcal{H}_{\rm eff} \delta_{\chi}' \simeq  4\pi G_{\rm eff} \rho_{\chi} \delta_{\chi} \mathcomma
    \label{eq:deltapp}
\end{equation}
where we have defined the effective Hubble term
\begin{equation}
   \mathcal{H}_{\rm eff} = \mathcal{H} \left(1 + \frac{Q}{\rho_{\chi}} \frac{\phi'}{\mathcal{H}}\right) \mathcomma
   \label{eq:heff}
\end{equation}
which involves an additional friction contribution related to the changes to the background expansion evolution, and the effective gravitational constant, which in the small-scale limit (large $k$) becomes dominant and is given simply as 
\begin{eqnarray}
G_{\rm eff} \simeq G_N\left(1 + 2 \text{M}_{\rm Pl}^2 \frac{Q^2}{\rho_{\chi}^2} \right) \mathcomma
\label{eq:geff}
\end{eqnarray}
as expected according to the general results for scalar-tensor gravity models under a conformal transformation \cite{Mifsud:2017fsy,Tsujikawa:2007gd,Amendola:2003wa}.

\section{Phenomenology} \label{sec:results}

In this Section, we explore the dynamics of DM and DE, discuss the main signatures left by the hybrid dark sector on the cosmological observables, and compare the predictions against $\Lambda$CDM. As expected, the qualitative features of the model are in line with standard coupled quintessence scenarios with constant effective interactions \cite{Wetterich:1994bg,Amendola:1999er} (see \cite{Barros:2018efl,Gomez-Valent:2020mqn,Teixeira:2019hil,Amendola:2000ub,Pettorino:2012ts,Pettorino:2013oxa,Xia:2013nua,vandeBruck:2016hpz,VanDeBruck:2017mua,Planck:2015bue,Planck:2018vyg,Agrawal:2019dlm} for recent studies). Nevertheless, there are distinct quantitative signatures due to the slow-rolling of the scalar field, on which we wish to focus. For illustration purposes, we consider four different realisations of the evolution of the model, characterised by $\phi_i/\text{M}_{\text{Pl}} = \{8,10,15,20\}$, with $\phi_i$ being the free parameter responsible for setting not only the initial dynamics of the scalar field but also the strength of the coupling in the dark sector. The initial velocity is kept constant at $\phi'_i = 0$ since it has no significant impact on the overall dynamics, as we will verify in more detail below. The cosmological parameters are fixed to standard \textit{Planck} 2018 fiducial values for a $\Lambda$CDM cosmology \cite{Planck:2018vyg}: $H_0 = 67.56$ km/s/Mpc for the Hubble parameter, and $\Omega_b h^2 = 0.022$ and $\Omega_c h^2 = 0.12$ for the relative energy density of the baryon and dark matter fluids, with $h=H_0/100$. The Friedmann constraint sets the scale of the potential. For the perturbative analysis, we assume Gaussian adiabatic initial conditions, a scalar power spectrum with an amplitude of curvature fluctuations $A_s = 2.215 \times 10^{-9}$ at the pivot scale $k_{\rm piv}=0.05$ Mpc$^{-1}$, and with spectral index $n_s = 0.962$. Moreover, and without loss of generality, we assume vanishing initial conditions for the scalar field perturbation and its corresponding velocity, that is, $\delta\phi_i = \delta\phi'_i = 0$. To calculate the evolution of the background and cosmological perturbations, we adapted the publicly available \texttt{CLASS} code\footnote{\href{https://github.com/lesgourg/class_public}{https://github.com/lesgourg/class\_public}} \cite{lesgourgues2011cosmic,Blas_2011,lesgourgues2011cosmic2} for our purposes. 

\subsection{Background evolution}

\begin{figure*}
      \subfloat{\includegraphics[height=0.36\linewidth]{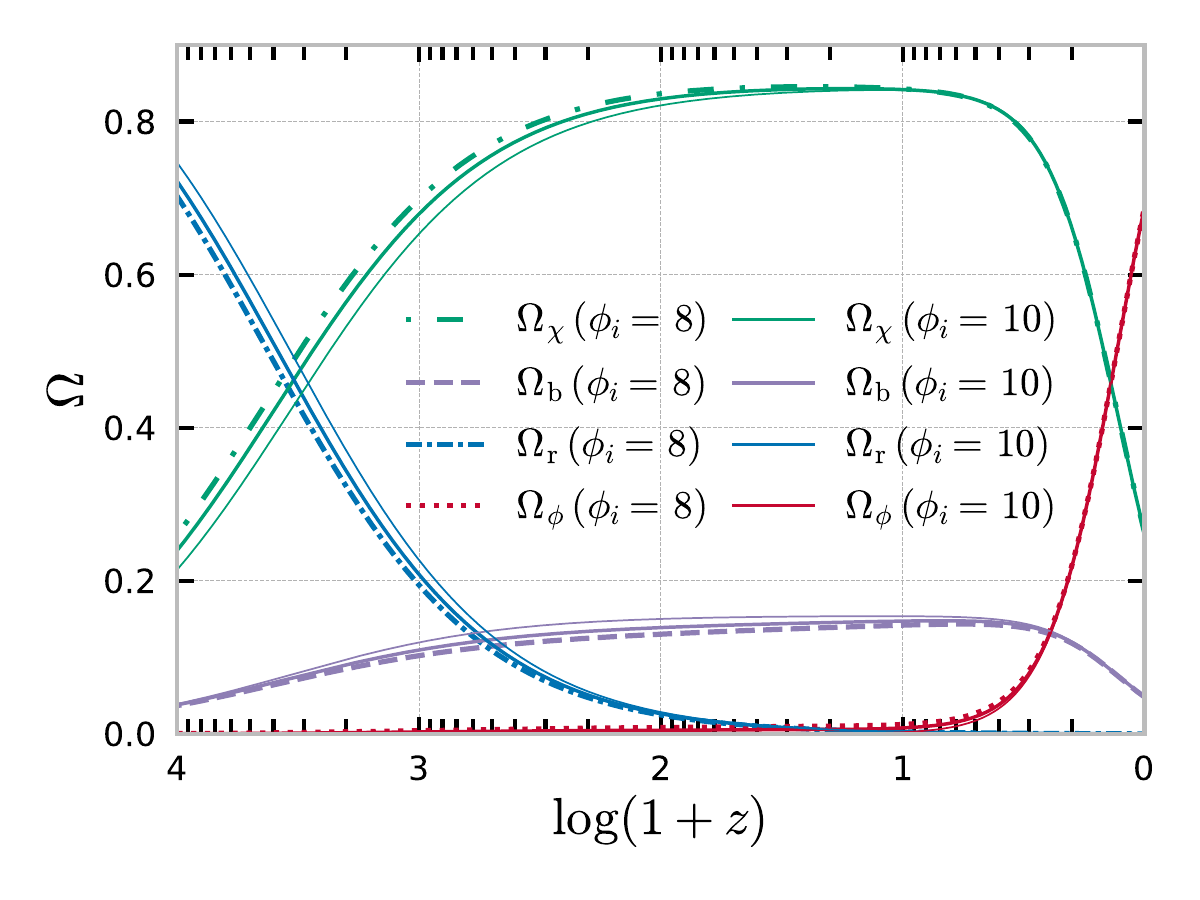}}
      \qquad
      \subfloat{\includegraphics[height=0.36\linewidth]{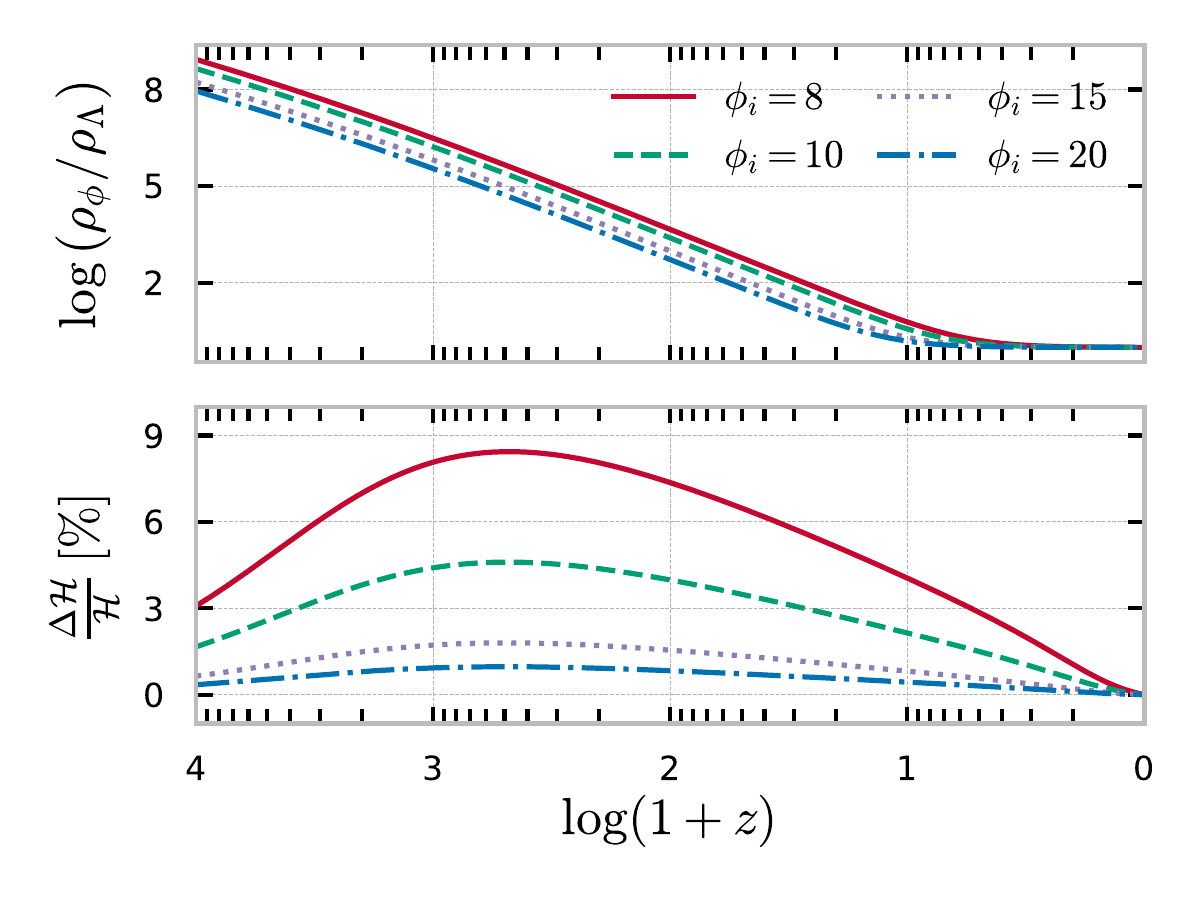}}
  \caption{\label{fig:oms} \textit{Left panel:} Redshift evolution of the relative energy densities $\Omega_i$ of the dark matter fluid $\chi$ (green), baryons (lavender), radiation (blue) and the scalar field $\phi$ (red) for the $\Lambda$CDM model (thin solid lines), $\phi_i = 8$ M$_{\text{Pl}}$ (dashed/dotted lines) and $\phi_i = 10$ M$_{\text{Pl}}$ (thick solid lines). \textit{Right panel:} Ratio of the dark energy density (\textit{top panel}) and fractional deviations in the Hubble rate (\textit{bottom panel}) in the hybrid coupled model with respect to the standard model as a function of redshift $1+z$, for $\phi_i = \{8,10,15,20\}$ M$_{\text{Pl}}$ (solid red, dashed green, dotted lavender and dot-dashed blue lines).}
\end{figure*}

\begin{figure*}
      \subfloat{\includegraphics[height=0.36\linewidth]{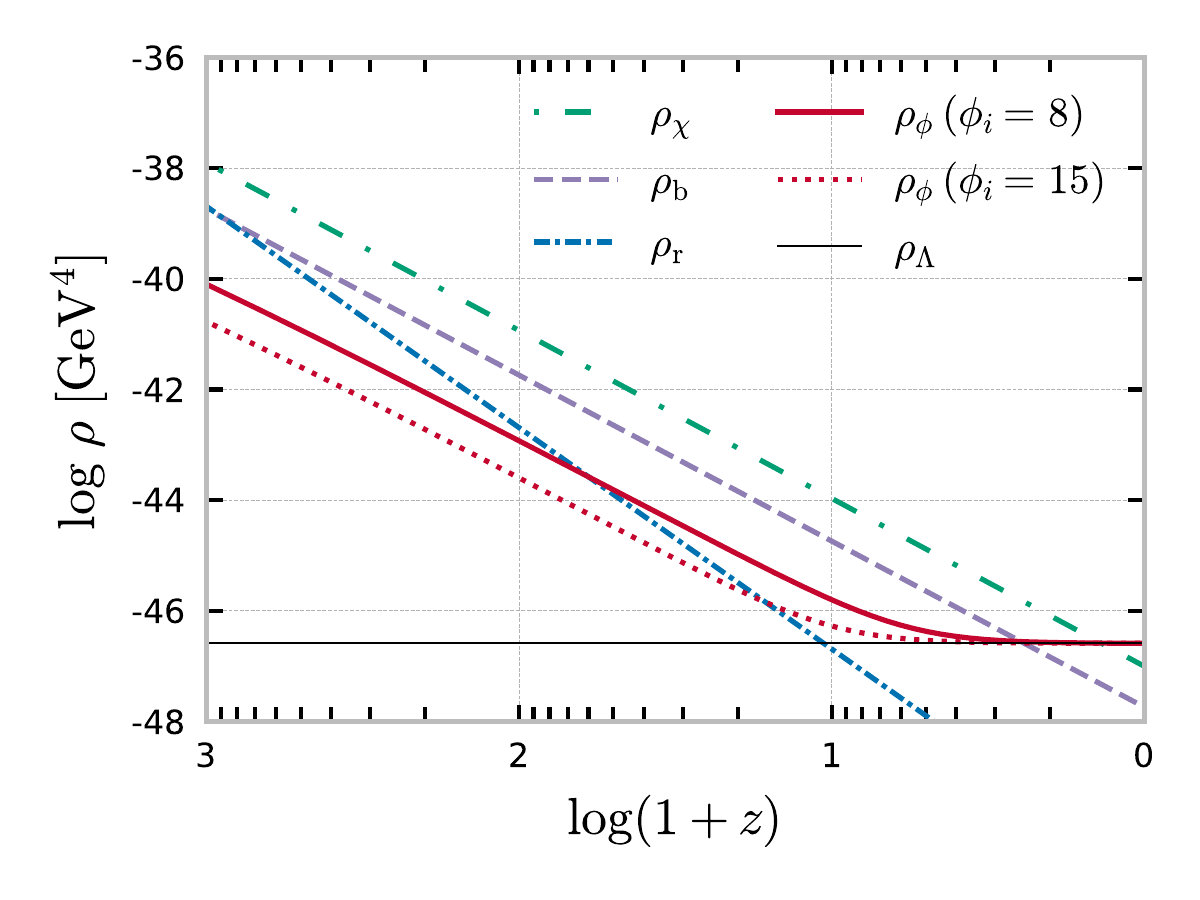}}
      \qquad
      \subfloat{\includegraphics[height=0.36\linewidth]{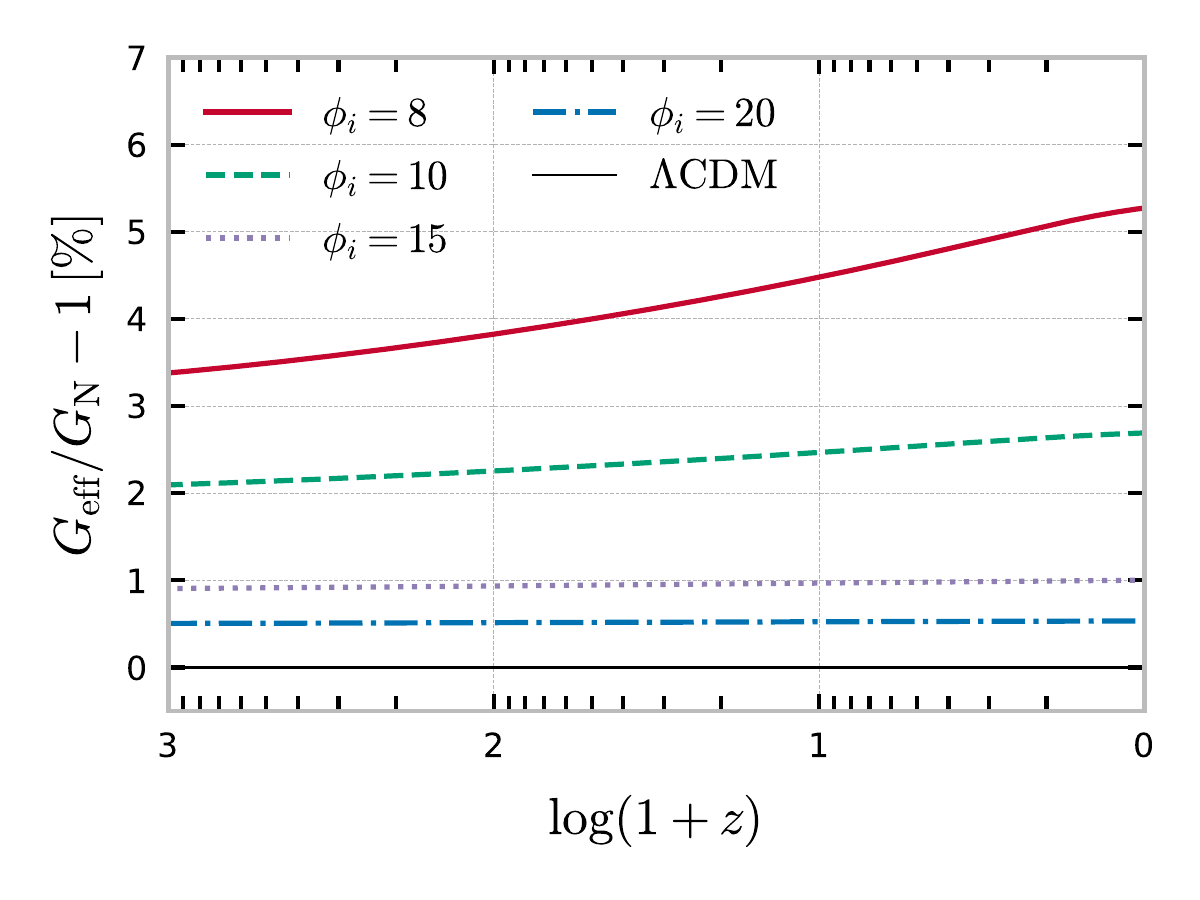}}
  \caption{\label{fig:back} \textit{Left panel:} Evolution of the energy densities $\rho$ of the dark matter fluid $\chi$ (green), baryons (lavender), radiation (blue) and the scalar field $\phi$ (filled red) for $\phi_i = 8$ M$_{\text{Pl}}$. To appreciate the differences, we also include $\rho_{\phi}$ for the $\phi_i = 15$ M$_{\text{Pl}}$ case (dotted red line) and $\rho_{\Lambda}$ for the standard model (thin black solid line) for completeness. \textit{Right panel:} Percentage deviations of the effective gravitational constant, as defined in Eq.~\eqref{eq:geff}, with respect to the standard $G_N$ (thin black solid line) for $\phi_i = \{8,10,15,20\}$ M$_{\text{Pl}}$ (solid red, dashed green, dotted lavender and dot-dashed blue lines).}
\end{figure*}

Since the potential $V(\phi,\chi)$ is simply constant in the fluid approximation, in the absence of the coupling, the field's evolution would be indistinguishable from a cosmological constant. It is the interaction between the scalar field with DM that drives the energy density of DE, as seen in the left panel of Fig.~\ref{fig:back}. When the coupling becomes relevant, at the end of the radiation-dominated epoch, $\phi$ is no longer static and starts evolving slowly until the dark energy density mimics the evolution of the component to which it couples, DM. During this period, the scalar field contributes to the dynamics as an {\it effective} pressureless fluid and the scaling period ends when the kinetic energy of the scalar field becomes comparable to the potential energy, set by $V_0$, \textit{i.e.} when $\phi'^2 \lesssim a^2 V_0$. Because the value of $\phi'$ is also set by the initial conditions for $\phi$ and $\rho_{\chi}$ (see Eq.~\ref{eq:phidotanalytical}), the end of the scaling regime is also determined by $\phi_i$: higher values of $\phi_i$ lead to late time dynamics closer to $\Lambda$CDM, and therefore $\phi$ leaves the scaling regime earlier (see right panel of Fig.~\ref{fig:oms} and left panel of Fig.~\ref{fig:back}); on the other hand, independently of its initial value, $\phi'$ is rapidly adjusted as the field is driven down the minimum of the effective potential, meaning that $\phi'_i$ regulates the onset of the scaling period only, which happens earlier for the highest initial velocities. The latter effect is insignificant for the background dynamics since it takes place early in the radiation-dominated epoch when the contribution of the scalar field is subdominant, and $\rho_\phi$ will be tracking DM as soon as $\rho_\chi$ dominates the evolution of the background. A similar effect has been identified in \cite{Pettorino:2008ez}, driven by a significant acceleration of the scalar field instead. When the $\phi-$field exits the matter-scaling regime, it heads towards a cosmological constant-like attractor solution, where it will remain diluting with the expansion until the fluid approximation breaks down.

Finally, we see that the effect of the coupling on $\phi$ is also manifested through an amplification of $\rho_\phi$ during this regime for the higher coupling cases (smaller $\phi_i$), which naturally leads to earlier matter-dark energy equality. This shift is an artefact of having fixed the present cosmology: DM is losing energy to DE, so it must be more abundant at early times to compensate for this effect. On the other hand, this is also entangled with a more significant contribution of the coupling to the scalar field dynamics through Eq.~\eqref{eq:kgphip}. There is also a shift in the matter-radiation equality towards earlier times for decreasing values of $\phi_i$, as shown in the left panel of Fig.~\ref{fig:oms}. 

In the lower right panel of Fig.~\ref{fig:oms}, we also present the relative deviations in the Hubble rate $H(z)$ for the hybrid coupled model in contrast with the $\Lambda$CDM case. Despite having fixed $H_0$ to the same present value, $H(z)$ is enhanced up to $9\%$ for the model with the lowest $\phi_i$ during the matter-dominated epoch while being negligible in radiation domination. This relates to the enhancement of $\rho_\phi$ and $\rho_\chi$ and will be essential to understand the growth of matter perturbations at different scales. We turn our attention to the evolution of cosmological perturbations in the following.

\subsection{Cosmological Perturbations}

\begin{figure*}
      \subfloat{\includegraphics[height=0.36\linewidth]{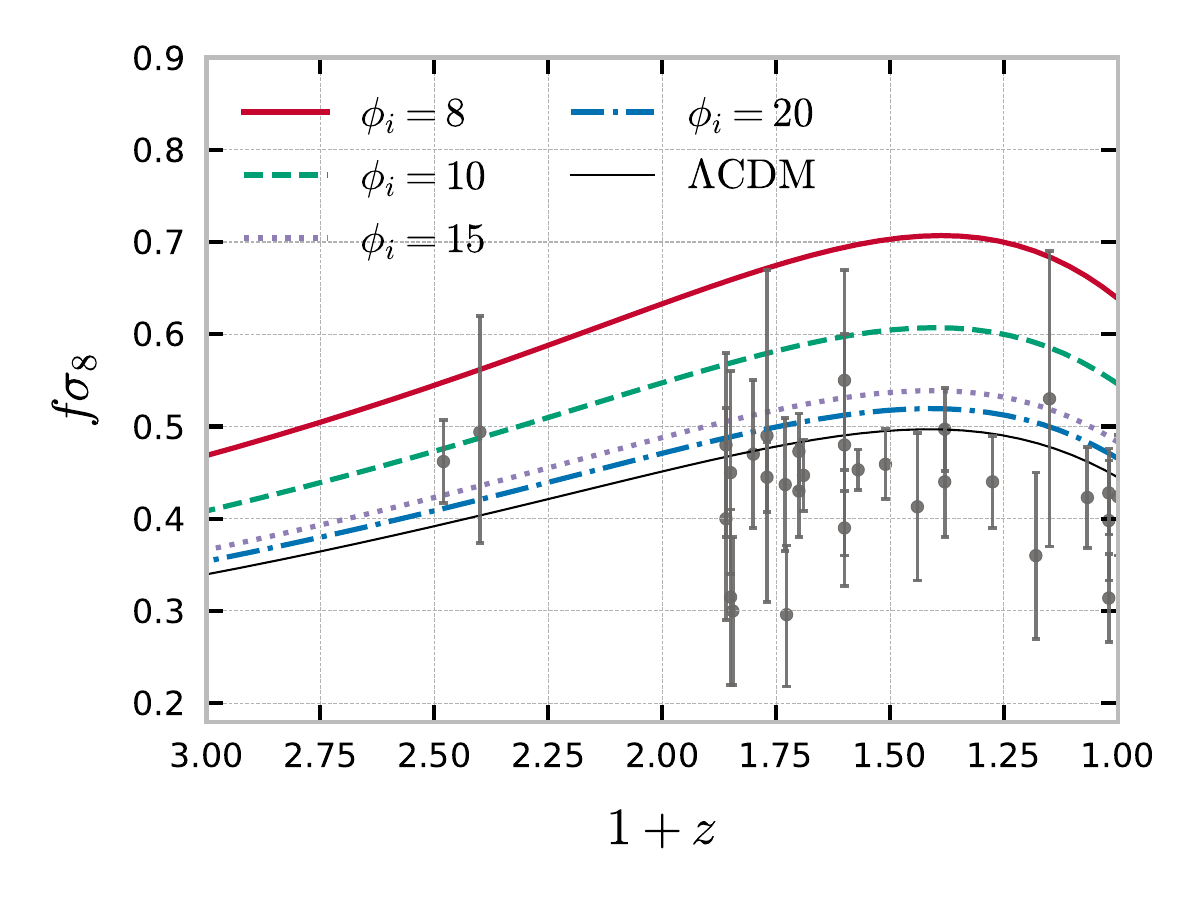}}
      \qquad
      \subfloat{\includegraphics[height=0.36\linewidth]{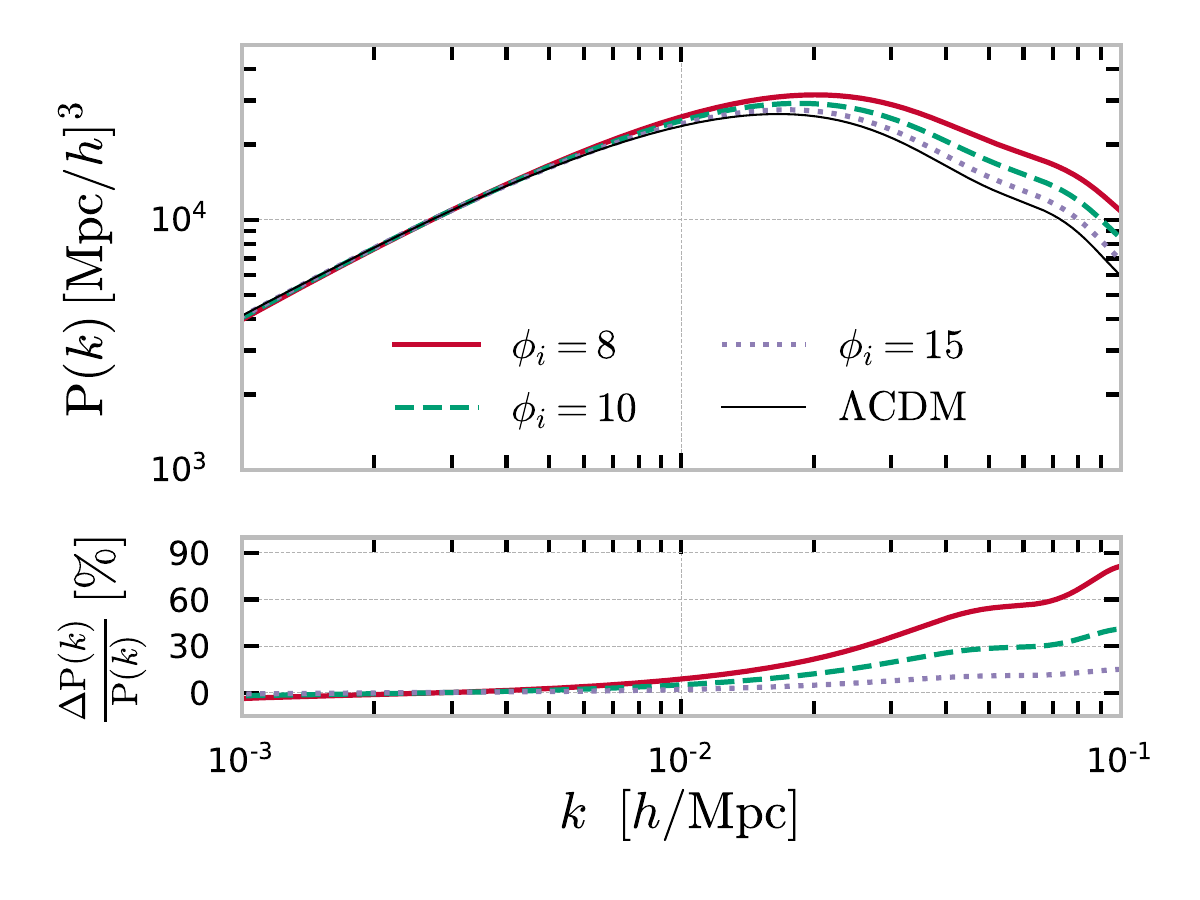}}
  \caption{\label{fig:growth} \textit{Left panel:} Evolution of the cosmological observable $f \sigma_8$ (defined in Eq.~\eqref{eq:fs8}) with redshift $1+z$ for the hybrid coupled model with $\phi_i = \{8,10,15,20\}$ M$_{\text{Pl}}$ (solid red, dashed green, dotted lavender and dot-dashed blue lines) and for the $\Lambda$CDM case (thin black solid line). The RSD data points and corresponding error bars (solid grey) correspond to the compilation presented in \cite{Marulli:2020uyy}. \textit{Right panel:} The matter power spectrum as a function of Fourier scales $k$ (\textit{top panel}) and corresponding percentage deviations (\textit{bottom panel}), for the hybrid coupled model with $\phi_i = \{8,10,15,20\}$ M$_{\text{Pl}}$ (solid red, dashed green and dotted lavender lines) with respect to the $\Lambda$CDM case (thin black solid line).}
\end{figure*}

The linear growth rate $f(a)$ of the total matter perturbation (\textit{i.e.} both baryons and DM) $\delta_m$ parametrises this effect and is defined by 

\begin{equation}
f (z,k) = \frac{1}{\mathcal{H}} \frac{\delta_m' (z,k)}{\delta_m (z,k)} \mathcomma
\end{equation}
where
\begin{equation}
\delta_m (z,k) = \frac{\Omega_b \delta_b + \Omega_{\rm DM} \delta_{\rm DM}}{\Omega_b + \Omega_{\rm DM}}. 
\end{equation}
The departure in the evolution of $f(z)$ in the hybrid model in contrast with $\Lambda$CDM coincides with the onset of the matter-dominated era when the coupling in the dark sector becomes important. The combined variable $f \sigma_8$ is directly connected to data since it is a scale-independent physical quantity that can be statistically constrained by observations of the growth of structures at different redshifts \cite{Linder:2016xer} and is expressed as
\begin{equation}
f \sigma_8 (z,k_{\sigma_8}) = \frac{\sigma_8 (0,k_{\sigma_8})}{\mathcal{H}} \frac{\delta_m' (z,k_{\sigma_8}) }{\delta_m (0,k_{\sigma_8}) } \mathcomma
\label{eq:fs8}
\end{equation}
where $\sigma_8$ is the root mean square mass fluctuation amplitude for spheres of size $8h^{-1}$ Mpc (or equivalently for scales $k_{\sigma_8} = 0.125 h$ Mpc$^{-1}$), parametrised as
\begin{equation}
 \sigma_8 (z,k_{\sigma_8}) = \sigma_8 (0,k_{\sigma_8}) \frac{\delta_m (z,k_{\sigma_8}) }{\delta_m (0,k_{\sigma_8})} \mathcomma
\end{equation}
 and it is generally used to set the amplitude of the matter power spectrum at present $\sigma_8^0 \equiv \sigma_8 (0,k_{\sigma_8})$. 

 Observations of redshift-space distortions (RSD) are a probe for the evolution of the matter perturbations. This effect arises from a Doppler shift ascribed to changes in the peculiar velocities of galaxies moving in clusters. Hence it can be used as a probe for the linear growth of structures and is observed as an additional contribution to the expansion redshift, with the redshift distribution of galaxies appearing ``distorted''. 
In the left panel of Fig.~\ref{fig:growth}, we present the redshift evolution of $f \sigma_8$ for the models considered, as defined in Eq.~\eqref{eq:fs8}. We observe an overall enhancement of the linear growth of matter perturbations, with the largest deviations from $\Lambda$CDM identified in the model with the lowest value of $\phi_i$, related to modifications in the expansion history. From the numerical study, we conclude that $\sigma_8^0$ is enhanced for all the models to achieve the same amplitude $\mathcal{A}_s$, with larger values of $\phi_i$ consistently approaching the $\Lambda$CDM curve. In Fig.~\ref{fig:growth} we also include observational data of RSD\footnote{\href{https://gitlab.com/federicomarulli/CosmoBolognaLib/tree/master/External/Data/}{https://gitlab.com/federicomarulli/CosmoBolognaLib}} from the compilation presented in \cite{Marulli:2020uyy} (see references therein) comprising measurements reported by various surveys. As one can see in the plots, in general the hybrid model predicts an enhancement of the linear growth, corresponding to a larger $f\sigma_8$, associated with higher values of $\sigma_8$ as well, assuming the same initial amplitude of primordial perturbations, $A_s$. Therefore, in these particular conditions, these models do not seem to provide a solution for the $S_8$--tension \cite{DiValentino:2020vvd,Abdalla:2022yfr}, although only a more thorough analysis could confirm this since the data--analysis implicitly assumes the $\Lambda$CDM model. 

In the right panel of Fig.~\ref{fig:growth}, we plot the power spectrum of matter density fluctuations $P(k)$ (top) and the corresponding relative deviations (bottom) for Fourier scales in the range $10^{-3}h\,\text{Mpc}^{-1} < k < 10^{-1}h\,\text{Mpc}^{-1}$. We find a slight suppression for the largest scales (lowest $k$) and a significant enhancement for intermediate and small scales (highest $k$), with the deviations being as high as $81\%$ for the smallest value of $\phi_i$ at $10^{-1}h\text{Mpc}^{-1}$ when the linear approximation starts to break down, and non-linear effects take over. This behaviour at large $k$ values is as expected since the fifth force affects the growth of perturbations and is more significant for smaller values of $\phi_i$.
 
The slight suppression at large scales (small $k$ values) can be explained by considering the modifications in the background evolution, in particular the deviation of the expansion rate $H(z)$, through the friction term in Eq.~\eqref{eq:heff}, inhibiting the growth of the matter density perturbations. This effect is dominant over the fifth-force at the largest scales only, reaching up to $4\%$ for the models considered, and is practically null for the models with the largest value of $\phi_i$, for which $G_{\rm eff} \sim G_N$, as shown in the right panel of Fig.~\ref{fig:back}. Moreover, we note that the turnover in the power spectrum is shifted towards smaller scales when compared with the $\Lambda$CDM case due to the change of the radiation-matter equality era towards larger redshifts.

\begin{figure*}
      \subfloat{\includegraphics[height=0.36\linewidth]{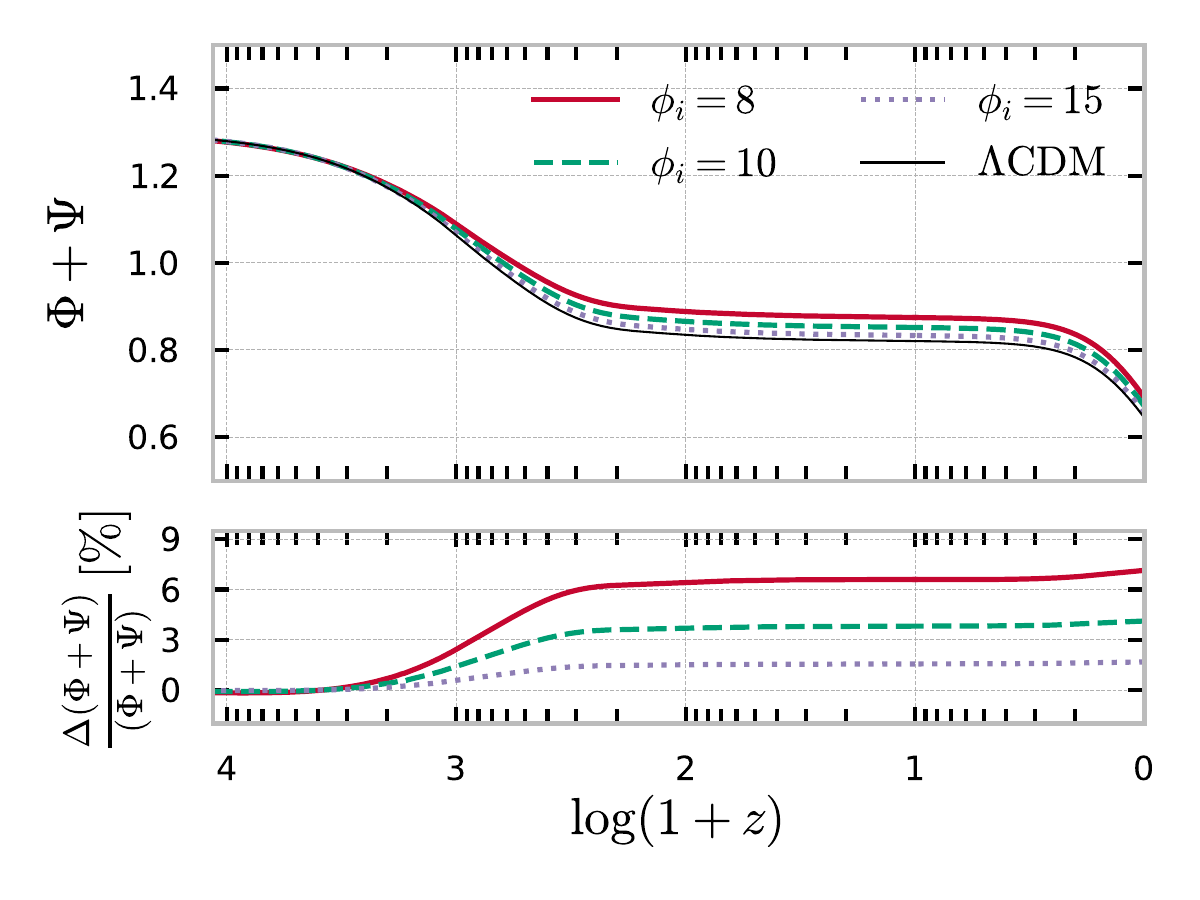}}
      \qquad
      \subfloat{\includegraphics[height=0.36\linewidth]{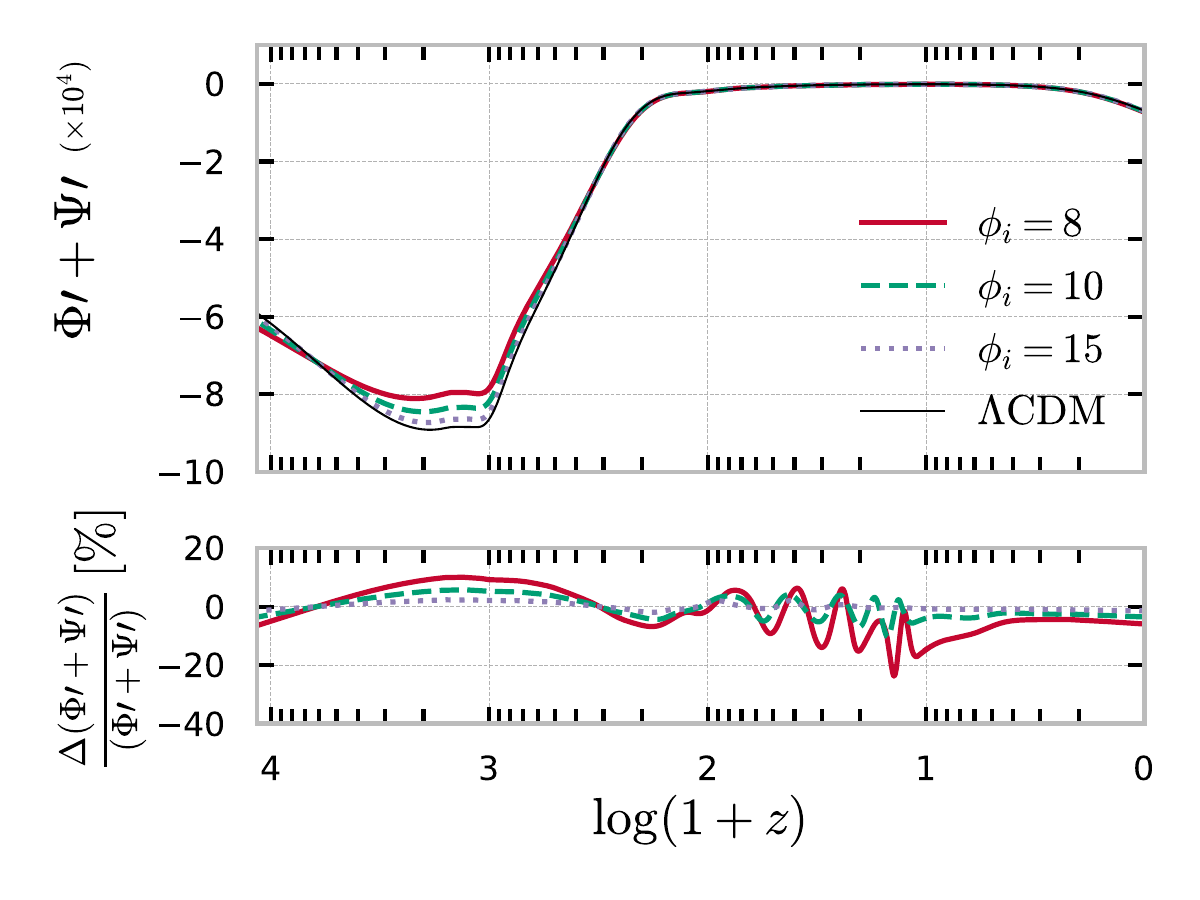}}
  \caption{\label{fig:lens} Redshift evolution of the sum of the gravitational potentials, $\Phi+\Psi$, (\textit{top left panel}) and the corresponding derivative with respect to conformal time, $\Phi'+\Psi'$ (\textit{top right panel}) for the hybrid coupled model with $\phi_i = \{8,10,15\}$ M$_{\text{Pl}}$ (solid red, dashed green and dotted lavender lines) and for the $\Lambda$CDM case (thin black solid line), including the percentage deviations from the standard model (\textit{bottom panels}). }
\end{figure*}

\begin{figure*}
      \subfloat{\includegraphics[height=0.36\linewidth]{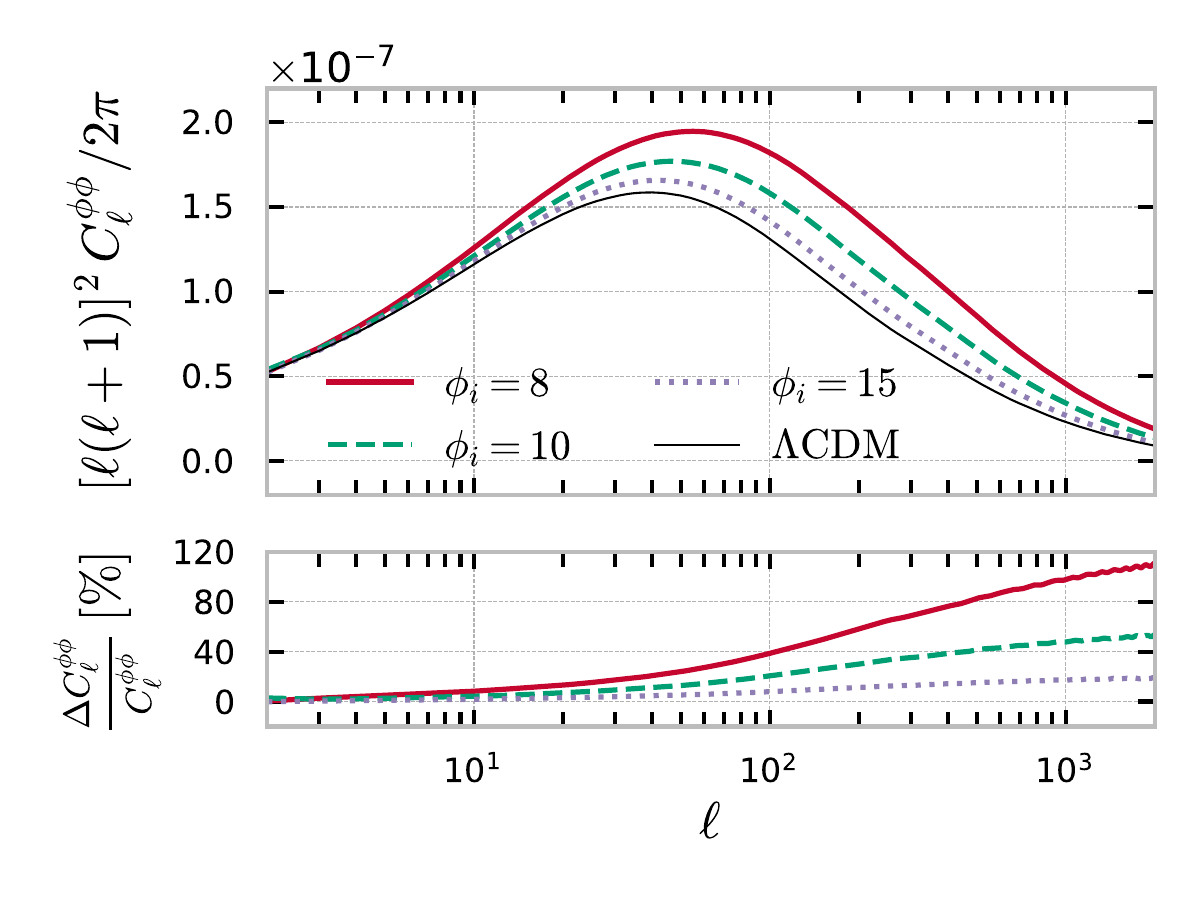}}
      \qquad
      \subfloat{\includegraphics[height=0.36\linewidth]{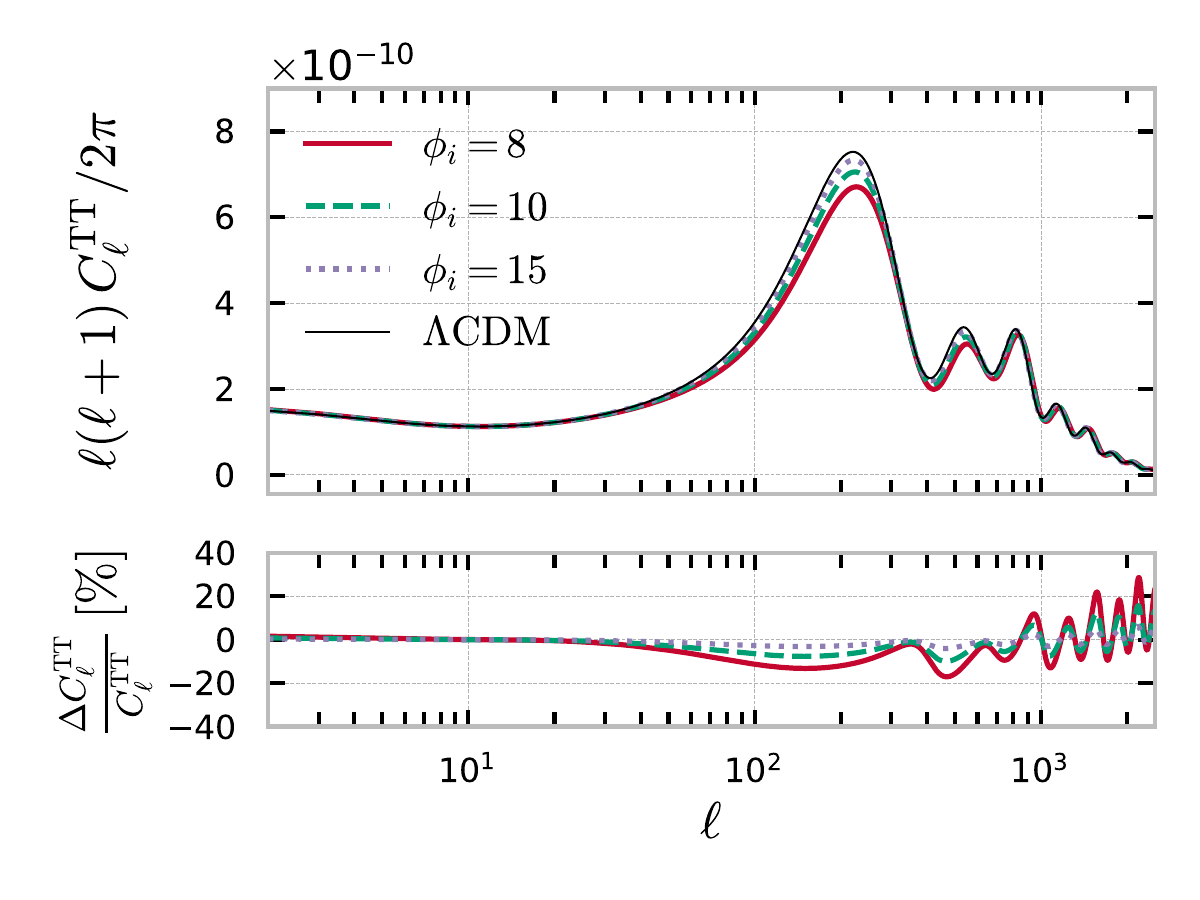}}
  \caption{\label{fig:cmb} Lensing (\textit{top left panel}) and TT (\textit{top right panel}) power spectra as a function of the angular scale $\ell$ for the hybrid coupled model with $\phi_i = \{8,10,15,20\}$ M$_{\text{Pl}}$ (solid red, dashed green, dotted lavender and dot-dashed blue lines) and for the $\Lambda$CDM case (thin black solid line), along with the fractional deviations from the standard model (\textit{bottom panels}).}
\end{figure*}

In the left panel of Fig.~\ref{fig:lens}, we depict the joint evolution of the gravitational potentials $\Phi$ and $\Psi$ (top) for the particular intermediate scale $k=0.01$ Mpc$^{-1}$ along with the percentage differences to $\Lambda$CDM. We observe that the most noticeable deviations occur for $z \lesssim 10^3$ during the matter-dominated era when the changes to the CDM evolution start to become relevant, and the scalar field is scaling with matter. As expected, we observe the most significant differences for the lowest values of $\phi_i$.
The modified evolution of the lensing gravitational potential $\phi_{len} = \Phi + \Psi$ can be attributed mainly to the exchange of energy from CDM to the DE field. The lensing potential is the relevant quantity in the source term for the line of sight integration in the computation of the lensing power spectrum $C_{\ell}^{\phi \phi}$, which we show in the left panel of Fig.~\ref{fig:cmb}. We find an overall amplification of $C_{\ell}^{\phi \phi}$ at all angular scales, with the highest deviations from $\Lambda$CDM for the lowest $\phi_i$ case. This enhancement could help accommodate the observed lensing excess in the \textit{Planck} temperature data of CMB anisotropies \cite{Planck:2018vyg,DiValentino:2019dzu}. Likewise, we can interpret this result according to the enhancement of the effective gravitational interaction of the DM particles. In fact, we observe that the matter density contrast $\delta_m$ follows the same trend over most of the Fourier scales considered, reflecting the increase of the effective gravitational constant, as shown in Fig.~\ref{fig:back}. 

The CMB anisotropies' temperature-temperature (TT) power spectrum, presented in the right panel of Fig.~\ref{fig:cmb}, encodes the same effects. The most significant contribution to the modifications in the TT power spectrum comes from the integrated-Sachs-Wolfe (ISW) effect, directly proportional to the time derivative of the lensing potential $\phi_{len}$ shown in the right panel of Fig~\ref{fig:lens}. We can split this effect into two different contributions: first, the early ISW effect results in an enhancement of the time derivative of the potentials as a direct consequence of an earlier transition from the radiation to the matter-dominated epochs; second, and most importantly, the late time ISW effect that arises due to changes in the lensing of the CMB by large scale structures induced by the non-trivial modified dynamics in the dark sector. The latter also results in a suppression of $\Phi' + \Psi'$ at late times, as depicted in the right panel of Fig.~\ref{fig:lens}. The most apparent modification in the TT power spectrum is the suppression of the amplitude of the peaks and troughs and the narrowing of their shapes associated with the decrease of the baryon to DM energy density ratio $\rho_{\rm b}/\rho_{\rm DM}$ during recombination. These are well-known effects studied in the literature \cite{Pettorino:2013oxa,Amendola:2011ie} that lead to a degeneracy between the effective coupling and the Hubble parameter since the latter mainly impacts the first peak's position and amplitude. 
In the same manner, the overall shift in the position of the acoustic peaks to higher multipoles is related to the changes in the expansion history, which modify the distance to the last scattering surface. Consequently, this results in a lower value for the sound horizon at the baryon-drag epoch (or, equivalently, a lower value for the angle subtended by the sound horizon at the decoupling time) when compared to $\Lambda$CDM. The more pronounced this shift is, the more drastic the deviations of the background evolution when compared to $\Lambda$CDM. Indeed we see from both panels of Fig.~\ref{fig:cmb} that the enhancement of the Hubble rate drives the CMB power spectra towards smaller angular scales (larger multipoles). Finally, the lensing power spectrum enhancement is associated with the enhancement of the ISW tail at large angular scales (low multipoles), although this effect is generally subdominant.

\section{Conclusions} \label{sec:conc}

In this work, we have proposed a hybrid model for the dark sector, in which DM and DE originate from two interacting scalar fields. We employed a form of potential commonly used in hybrid inflation to model the DM-- DE system. The cosmology in this setup is studied in considerable detail. The heavy scalar field quickly oscillates from deep inside the radiation-dominated epoch and consequently behaves like pressureless DM. We have shown that, once the heavy field starts to oscillate rapidly, the two scalar fields can be described by a DM fluid coupled to a slowly evolving DE field. 

In closing we highlight the following predictions of the model proposed: 

\begin{itemize}

\item We find that the DE field must have a large field value today ($\phi> \text{M}_{\text{Pl}}$) so that the fluid description is valid. Consequently, the coupling between DM and DE is relatively small today, and the DM is very heavy (similar to DM in the WIMPZilla scenario). At the same time, the energy scales in the potential are significantly reduced compared to the Planck scale. The mass scale $M$ is of order eV or so, depending on the coupling constant $\lambda$. The fact that this model requires super-Planckian field excursions, like in inflationary scenarios, provides a challenge to model building in theories beyond the standard model. But it is interesting to note that the model proposed here combines two mass scales: the small mass scale $M$ and the large field excursions for the DE field $\phi$. 

\item Another prediction of the model is that the epoch of dark energy domination is transient. In future, the DM field becomes light and will no longer behave like a pressureless fluid. Both scalar fields will settle at the true minimum of the potential (at $\phi = 0$ and $\chi = \pm M$). The Universe's future will then depend on whether space is closed. If the Universe is closed, the expansion will stop and collapse, opening up the possibility of a bounce in the long-distant future. 
\end{itemize}

As we have seen, if the coupling is large enough today, the DE field leaves distinct signatures on the temperature-temperature power spectrum of CMB anisotropies and in the growth of structures encoded in the matter power spectrum. These characteristic changes can be tested against current and future observational data. A more thorough and detailed study is needed, resorting to different independent data sets and sampling the remaining relevant cosmological parameters. We leave this for future work. 

\acknowledgments

CvdB is supported (in part) by the Lancaster–Manchester–Sheffield Consortium for Fundamental Physics under STFC grant: ST/T001038/1. G.P. is supported by a EPSRC studentship. E.M.T. is supported by the grant SFRH/BD/143231/2019 from Funda\c{c}\~ao para a Ci\^encia e a Tecnologia (FCT). This article is based upon work from COST Action CA21136 Addressing observational tensions in cosmology with systematics and fundamental physics (CosmoVerse) supported by COST (European Cooperation in Science and Technology). 

\appendix


\bibliography{bibliography}

\end{document}